\documentclass[twocolumn,
11pt,tightenlines,nofootinbib,preprintnumbers,showpacs,floatfix,prd]{revtex4}
\usepackage{graphicx}
\usepackage{amsmath}
\input epsf

\newcommand{\mbf}{\mathbf}
\begin{document}
%\preprint{CPT-PXX-2007}
\title{Impact picture for the analyzing power $A_N$ 
in very forward $pp$ elastic scattering}
%\title{The analyzing power $A_N$ for p p elastic scattering in the 
%very forward direction in the impact-picture model}

\author{Claude Bourrely\footnote{Electronic address: 
Claude.Bourrely@cpt.univ-mrs.fr}}
\affiliation{Centre de Physique Th\'eorique\footnote{Unit\'e Mixte 
de  Recherche 6207 du CNRS et des Universit\'es Aix-Marseille I,
Aix-Marseille II et de l'Universit\'e du Sud Toulon-Var -
Laboratoire affili\'e \`a la FRUMAM.}, 
CNRS Luminy case 907,\\
F-13288 Marseille Cedex 09, France}
\author{Jacques Soffer\footnote{Electronic address: jsoffer@temple.edu}}
\affiliation{Department of Physics, Temple University,
Philadelphia, PA 19122-6082, USA}
\author{Tai Tsun Wu\footnote{Electronic address: ttwu@seas.harvard.edu}}
\affiliation{Harvard University, Cambridge, MA 02138, USA and\\
Theoretical Physics Division, CERN, 1211 Geneva 23, Switzerland}
\begin{abstract}
In the framework of the impact picture we compute the analyzing power $A_N$
for $pp$ elastic scattering at high energy and in the very forward direction. We
consider the full set of Coulomb amplitudes and show that
the interference between the hadronic non-flip
amplitude and the single-flip Coulomb amplitude is sufficient to obtain
a good agreement with the present experimental data. This leads us to conclude
that the single-flip hadronic amplitude is small in this low momentum
transfer region and it strongly suggests that this process can be used as an absolute
polarimeter at the BNL-RHIC $pp$ collider.
\end{abstract}
\pacs{ 13.40.-f, 13.85.Dz, 13.88.+e}
\maketitle
%%%%%%%%%%%%%%%%%%%%%%%%%%%%%%%%%%%%%%%
\section{Introduction}
The measurement of spin observables in hadronic exclusive processes is the only
way to obtain the full knowledge on the corresponding set of scattering amplitudes, 
and in particular, their relative size and phase difference. Taking the specific
case of proton-proton elastic scattering, a reconstruction of the five
amplitudes has been worked out in the low-energy domain \cite{leluc}. This situation
is very different at high energy; due to the lack of data, in the range $p_{lab} \simeq$ 100-300 GeV,
besides the non-flip hadronic amplitude $\phi_1^h$, only the hadronic helicity-flip amplitude 
$\phi_5^h$ is known and to a rather poor level of accuracy. 
The advent of the
BNL-RHIC $pp$ collider, where the two proton beams can be polarized, longitudinally and 
transversely, up to an energy
$\sqrt{s} = 500~\mbox{GeV}$, offers a unique opportunity to measure
single- and double-spin observables, and thus to provide the determination
of the spin-dependent amplitudes, which remain unknown so far.

For instance, for an elastic collision of transversely polarized protons,
the differential cross section as a function of the momentum
transfer $t$ and the azimuthal angle $\phi$, reads
\begin{align}
2\pi\frac{d^2\sigma}{dt d\phi} =& \frac{d\sigma}{dt}
[1 + (P_B + P_Y) A_N \cos\phi \\
\nonumber
& +P_BP_Y (A_{NN}\cos^2\phi + A_{SS}\sin^2\phi)]~,
\label{anexp}
\end{align}
where $P_B$ and $P_Y$ are the beam polarizations, $A_N$ the analyzing power
and $A_{NN}$, $A_{SS}$
are double spin asymmetries (see Ref.~\cite{bou1} for definitions).
In this expression, the values of the beam polarizations have to be known 
accurately in order to reduce the errors on the spin asymmetries. So new measurements
are indeed required to achieve an amplitude analysis of $pp$ elastic scattering at high energy,
and the success of the vast BNL-RHIC spin programme \cite{bssv} also relies heavily on
the precise determination of the beam polarizations. One possibility for 
an absolute polarimeter\footnote{Proton-Helium elastic scattering has been also considered
as a possible high-energy polarimeter \cite{bou7}.} is provided by
the measurement of the analyzing power $A_N$, in the very forward $|t|$
region, where significant Coulomb nuclear interference (CNI) occurs \cite{bou6,bklst,butt01}.

In the calculation of the analyzing power an important question arises: is the
interference fully dominated by the hadronic non-flip amplitude with the 
one-photon exchange helicity-flip amplitude or must one also take into account
the contribution of the hadronic helicity-flip amplitude $\phi_5^h$, mentioned above? 
Several arguments concerning the magnitude and phase
of $\phi_5^h$ in the small $t$-region, have been discussed in great detail in Ref.~\cite{bklst}
and it was concluded that the measurement of $A_N$ in the CNI region was badly needed to
get the answer. The purpose of this paper is to study this problem in the framework
of the impact picture developed almost three decades ago \cite{bou3}, which has led to
a very successful phenomenology, repeatedly verified by high-energy experiments, 
including near the forward direction.\footnote{ An accurate measurement for the
real part of the $pp$ forward scattering amplitude is a real challenge for the LHC \cite{bkmsw}.} 
%%%%%%%%%%%%%%%%%%%%%%%%%%%%%%%%%%%%%%
\section{\bf The impact-picture approach}
\label{des}
In the impact picture, the spin-independent hadronic 
amplitude $\phi_1^h=\phi_3^h$
for $p p$ and $\bar p p$ elastic scattering reads as \cite{bou3}
\begin{equation}\label{ampli}
\phi_{1,3}^h(s,t) = \frac{is}{2\pi}\int e^{-i\mbf{q}\cdot\mbf{b}} (1 - 
e^{-\Omega_0(s,\mbf{b})})  d\mbf{b} \ ,
\end{equation}
where $\mbf q$ is the momentum transfer ($t={-\bf q}^2$) and 
$\Omega_0(s,\mbf{b})$ is the opaqueness at impact parameter 
$\mbf b$ and at a given energy $s$. We take 
\begin{equation}\label{opac}
\Omega_0(s,\mbf{b}) = S_0(s)F(\mbf{b}^2)+ R_0(s,\mbf{b})~.
\end{equation}
Here the first term is associated with the Pomeron exchange, which generates 
the diffractive component of the scattering and the second term is 
the Regge background.
The Pomeron energy dependence is given by the crossing symmetric expression 
\cite{r1,r2} 
\begin{equation}\label{energ}
S_0(s) = \frac{s^c}{ (\ln s)^{c'}} + \frac{u^c}{ (\ln u)^{c'}} \ ,
\end{equation}
where $u$ is the third Mandelstam variable.
The choice one makes for $F(\mbf{b}^2)$ is crucial and, as explained in Ref.~\cite{bou3},
we take the Bessel transform of
\begin{equation}\label{formf}
\tilde F(t) = f[G(t)]^2~{a^2 + t \over a^2 -t} ~.
\end{equation}
Here $G(t)$ stands for the proton 
electromagnetic form factor, parametrized as 
\begin{equation}\label{fgt}
G(t) = {1 \over (1 - t/m_1^2)(1 - t/m_2^2)} \ .
\end{equation}
The slowly varying function occurring in Eq.~(\ref{formf}) reflects the 
approximate proportionality between the charge density and the hadronic matter
distribution inside a proton \cite{r3}.
So the Pomeron part of the amplitude depends on only {\it six} parameters
$c, c', m_1, m_2, f,$ and $a$. 
The asymptotic energy regime of hadronic interactions are controlled by 
$c$ and $c'$, which will be kept, for all elastic reactions, at
the values obtained in 1984 \cite{bou9}, namely
\begin{equation}\label{cc'}
c=0.167 ~~~\mbox{and}~~~c'=0.748~~.
\end{equation}
The remaining four parameters are related, more specifically to the reaction 
$pp$ ($\bar p p$) and they have been fitted in \cite{bou8} by
the use of a large set of elastic data.

We now turn to the Regge background. A generic Regge exchange amplitude 
has an expression of the form
\begin{equation}\label{ampreg}
\tilde R_i(s,t)=C_ie^{b_it} \left[ 1 \pm e^{-i\pi\alpha_i(t)}\right]
[\frac{s}{s_0}]^{\alpha_i(t)} \ ,
\end{equation}
where $C_ie^{b_it}$ is the Regge residue, $\pm$ refers to an even- or odd-signature exchange, 
$\alpha_i(t) = \alpha_{0i} + \alpha_i^{'} t$, is 
a standard linear Regge trajectory and $s_0 =1~$GeV$^2$.
If $\tilde R_0(s,t)= \sum_i \tilde R_i(s,t)$ is the sum over all the allowed Regge trajectories, 
the Regge background 
$R_0(s,\mbf{b})$ in Eq.~(\ref{opac}) is the Bessel transform of 
$\tilde R_0(s,t)$. In $pp$ ($\bar p p$) elastic scattering,
the allowed Regge exchanges are $A_2$, $\rho$, $\omega$, so the Regge 
background involves several additional parameters, which are given in
Ref.~\cite{bou8}.

In earlier work, spin-dependent hadronic amplitudes were implemented
\cite{bou3,bou4,bou5}, using the notion of rotating matter inside
the proton, which allowed us to describe the polarizations and spin correlation 
parameters, but for the present purpose hadronic spin-dependent amplitudes
will be ignored.
In order to describe the very small $t$-region we are interested in,
one adds to the hadronic amplitude considered above, the full
set of Coulomb amplitudes $\phi_i^C(s,t)$, whose expressions are given 
in Ref.~\cite{lap78} and the Coulomb phase in Ref.~\cite{kund05}.

The two observables of interest are the unpolarized cross section
$d\sigma/dt$ and the analyzing power $A_N$, whose expressions in terms
of the hadronic and Coulomb amplitudes are respectively
\begin{equation}
\frac{d\sigma(s,t)}{dt}\!\! =\!\! \frac{\pi}{s^2}
\sum_{i=1,\cdots, 5}\,|\phi^h_i(s,t) + \phi^C_i(s,t)|^2
\label{cross} 
\end{equation}  
and  
\begin{equation} 
 A_N(s,t)\!\! = \!\! \frac{4\mbox{Im}((\phi_1^h(s,t))^*\,\phi_5^C(s,t))}
{\sum_{i=1,\cdots, 5}\!|\phi^h_i(s,t)\! +\! \phi^C_i(s,t)|^2}~. 
\label{anth}
\end{equation}
The numerator of this last expression is not fully general because
we have assumed that $\phi_1^h=\phi_3^h$ and $\phi_{2,4,5}^h=0$.

\begin{figure}[thp]
       \vspace*{-38mm}
\begin{center}
     \begin{minipage}[t]{0.250\textwidth}
       \hspace*{-0.5\textwidth}
       \includegraphics[width=1.9\textwidth]{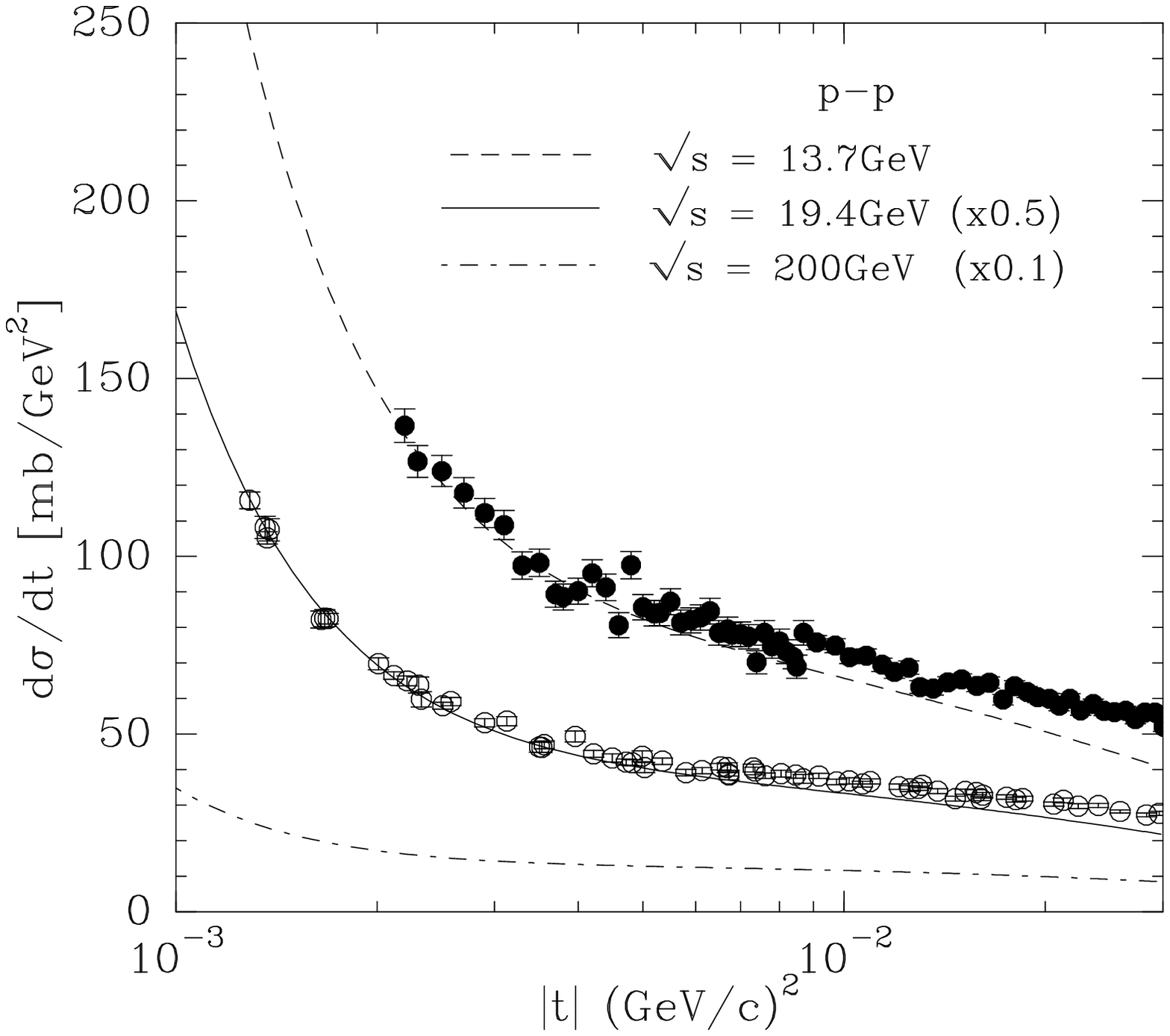}
       \vspace*{-65mm}
    \end{minipage}\hfill
\end{center}
      \vspace*{-0.2\textwidth}%
\begin{center}
    \begin{minipage}[t]{0.250\textwidth}
      \hspace*{-0.5\textwidth}
      \includegraphics[width=1.9\textwidth]{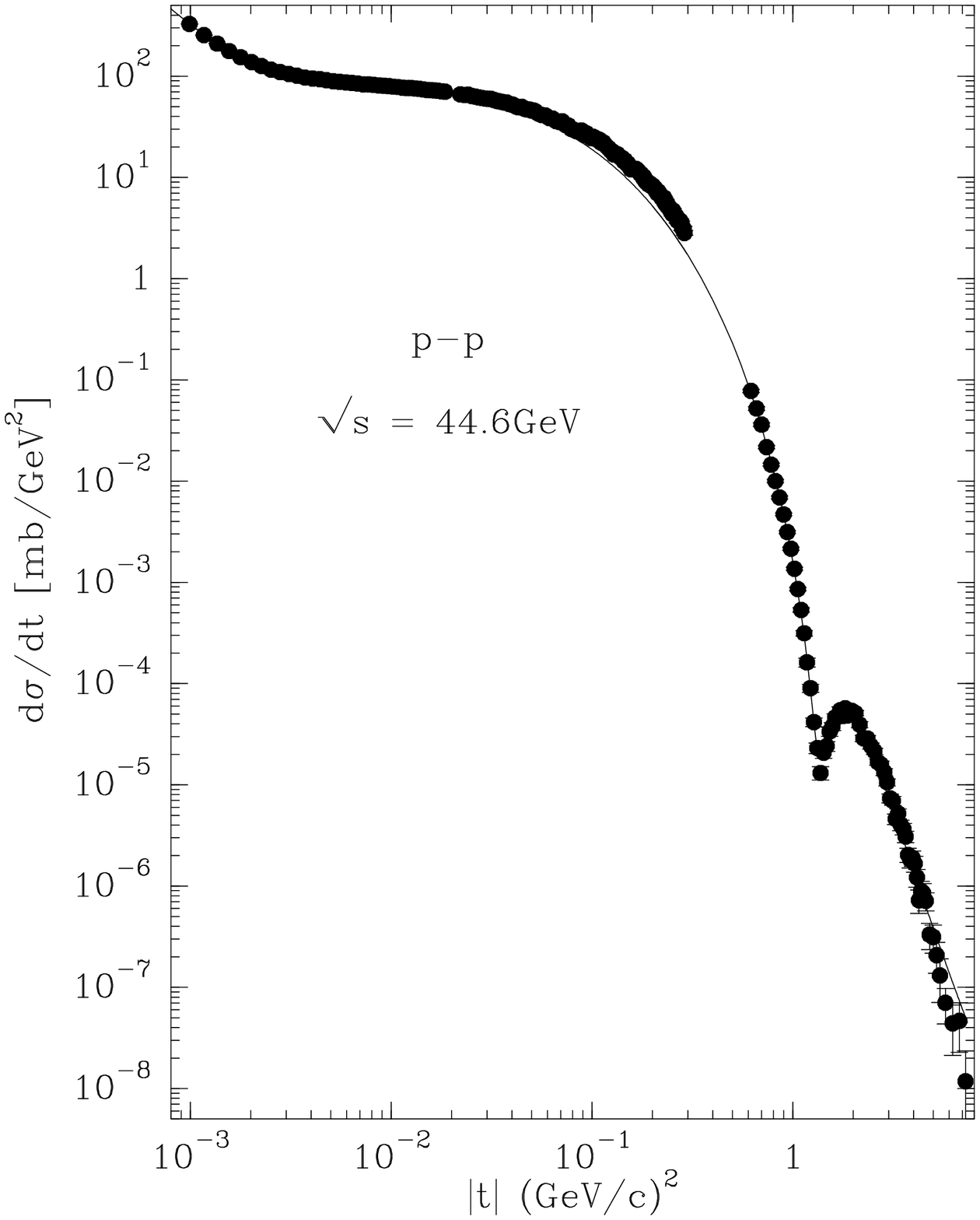}
       \vspace*{-65mm}
    \end{minipage}\hfill
\end{center}
\vspace*{-19mm}
\caption{\label{fig:1}The differential cross section versus the momentum
transfer $t$ for different energies.
Data from Refs.~ \cite{bart72,schiz81,burq83,fajardo,gros78,amal79}.}
\vspace*{-1.2ex}
\end{figure}
%%%%%%%%%%%%%%%%%%%%%%%
\begin{figure}[thp]
       \vspace*{-10mm}
\begin{center}
     \begin{minipage}[t]{0.250\textwidth}
       \hspace*{-0.5\textwidth}
      \includegraphics[height=2.3\textwidth]{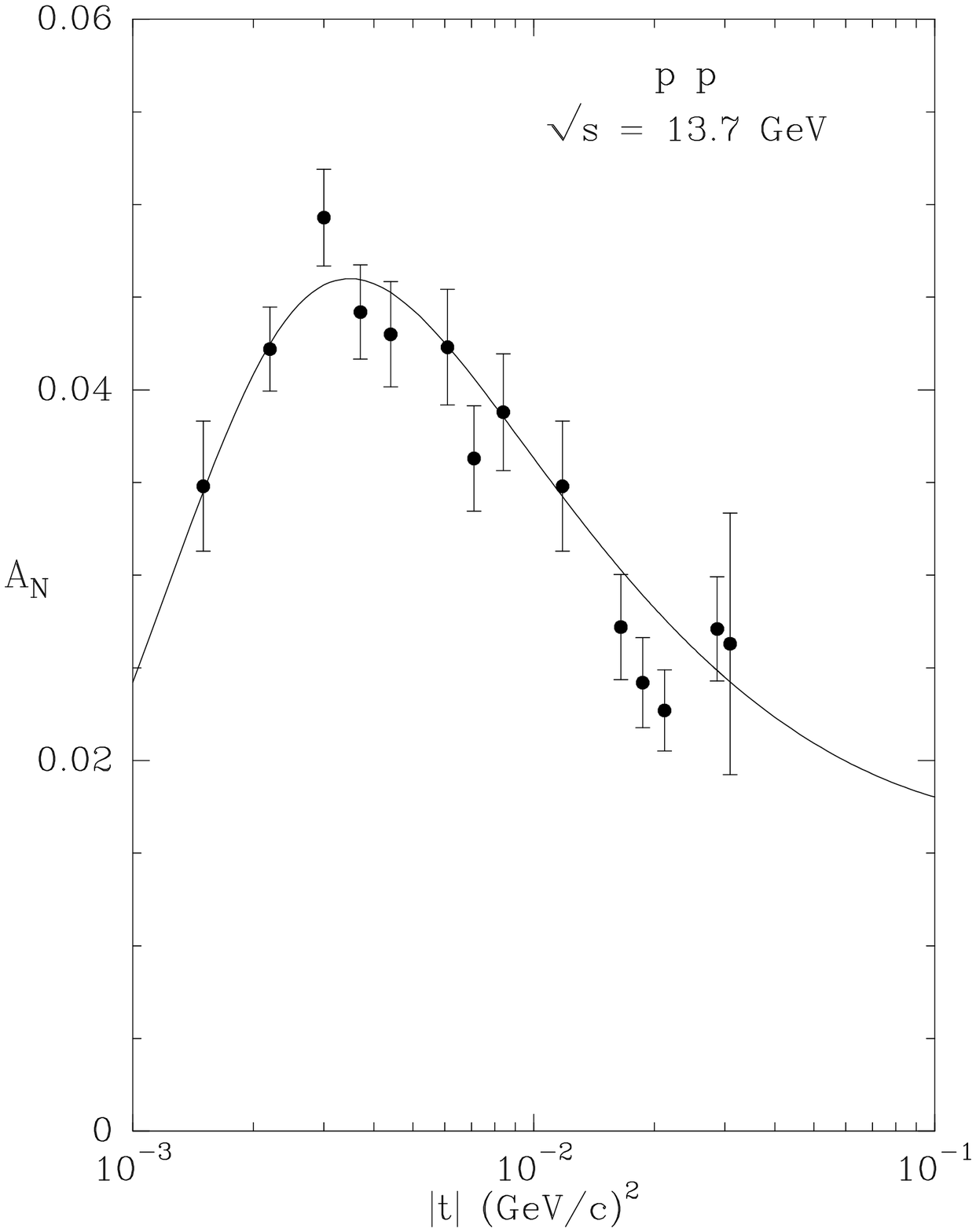}
       \vspace*{-60mm}
    \end{minipage}\hfill
\end{center}
      \vspace*{-0.2\textwidth}%
\begin{center}
    \begin{minipage}[t]{0.250\textwidth}
      \hspace*{-0.5\textwidth}
      \includegraphics[height=2.3\textwidth]{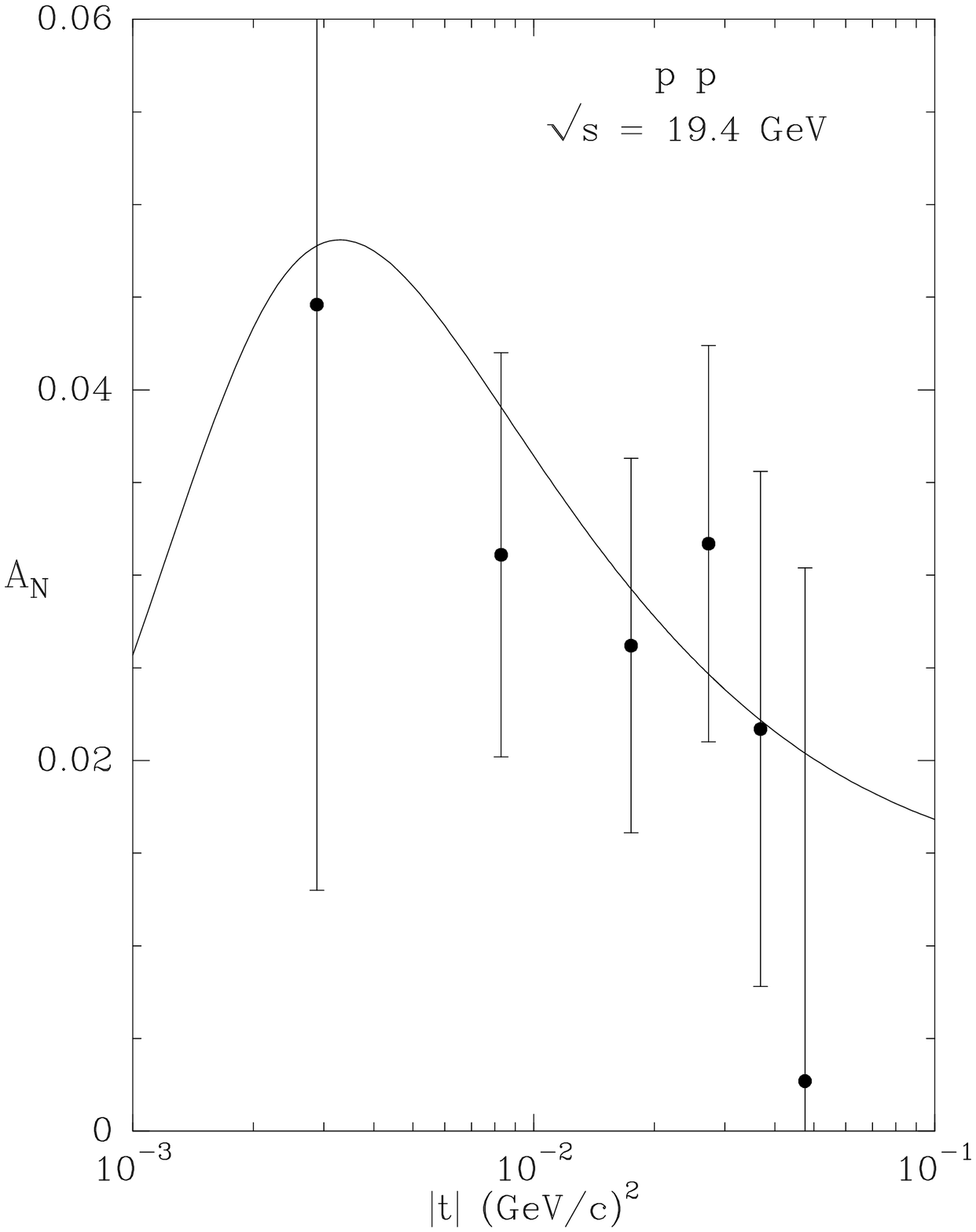}
       \vspace*{-60mm}
    \end{minipage}\hfill
\end{center}
      \vspace*{-0.2\textwidth}%
\begin{center}
    \begin{minipage}[t]{0.250\textwidth}
      \hspace*{-0.5\textwidth}
      \includegraphics[height=2.3\textwidth]{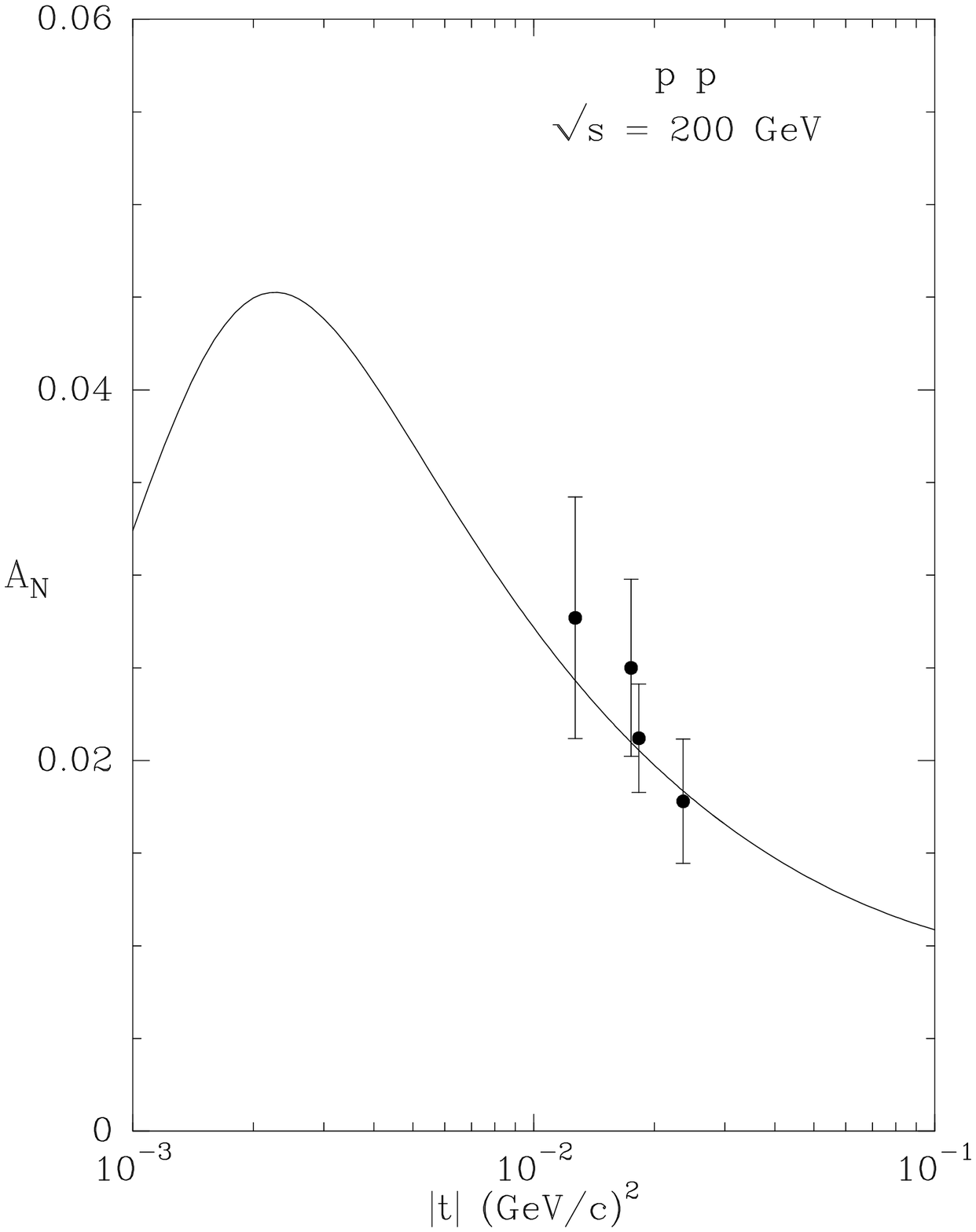}
       \vspace*{-60mm}
    \end{minipage}\hfill
\end{center}
\vspace*{-16mm}
\caption{\label{fig:2}The analyzing power $A_N$ versus the momentum
transfer $t$ for different energies. 
Data from Refs.~ \cite{okada05,akch93,bult06}.}
\vspace*{-1.2ex}
\end{figure}
%%%%%%%%%%%%%%%%%%%%%%%%%%%%%%%%%%%%%%
\section{Numerical results}
\label{secan}
The analyzing power $A_N$ has been measured at high energy for
$\sqrt{s} = 13.7,~19.4,~200~ \mbox{GeV}$, but before turning to the calculation of this quantity,
it is necessary to look at the predictions for the differential
cross section, at the corresponding energies. They are given in the upper plot in 
Fig.~\ref{fig:1} and compared with the available experimental results at $\sqrt{s}$ = 13.7 and 19.4~GeV.
We underestimate a bit the data for high $t$-values, at $\sqrt{s} = 13.7~\mbox{GeV}$, which might indicate
the presence of a small hadronic spin-dependent amplitude. However, this is not the case at
$\sqrt{s} = 44~\mbox{GeV}$, where the agreement is excellent, as shown in the lower plot in Fig.~\ref{fig:1}.
Note that the momentum transfer runs over four decades and the cross section over eleven orders
of magnitude, which is a good illustration of the validity of the impact picture. Concerning the energy
$\sqrt{s} = 200~\mbox{GeV}$, we cannot make a detailed comparison with the data. The pp2pp experiment
\cite{bult04} has only determined the slope of the cross section for $0.01<|t|<0.019 ~\mbox{GeV}^2$, which
is $b=16.3 \pm 1.6 (stat.) \pm 0.9 (syst.)~ \mbox{GeV}^{-2}$, consistent with the average value
obtained in the impact picture, namely $b=16.25 ~\mbox{GeV}^{-2}$.

\begin{figure}[thp]
\includegraphics[width=8cm]{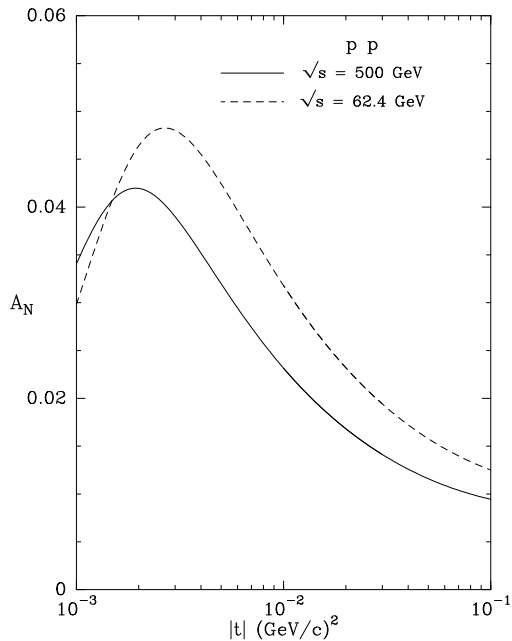}
\caption{\label{fig:3}Predictions for the analyzing power $A_N$ versus the momentum
transfer $t$ for two energies.}
\end{figure}

In Fig.~\ref{fig:2}, we compare the predictions with the data, for $A_N$ in the CNI region
versus $|t|$, for three different energies and let us make the following 
remarks. First, there is almost no energy dependence between $\sqrt{s}$ = 13.7 and 19.4 GeV, but
the curve has a slightly different shape at $\sqrt{s}$ = 200 GeV. Second,
although this is not obvious from the plot, $A_N$ does not vanish for $|t|>0.1~\mbox{GeV}^2$ and
we would like to stress that for $pp$ elastic scattering at high energy,
in the dip region, the hadronic and the Coulomb amplitudes are of the same
order of magnitude \cite{bou2}, so the behavior of spin observables
is sensitive to this interference. Finally, indeed
the predictions agree well with the present experimental
data and in view of future data taking in the BNL-RHIC spin
programme, we display in Fig.~\ref{fig:3} some predictions at $\sqrt{s}$ = 62.4 GeV and $\sqrt{s}$ = 500 GeV.
When the energy increases the maximum of $A_N$ decreases and occurs at a lower $t$ value, which clearly reflects
the rise of the total cross section \cite{bklst}.
 The above discussion shows that the hadronic spin-flip amplitude is not necessary to describe
the analyzing power, at least when compared with the data with the present day accuracy. A 
similar conclusion was obtained in Ref.~\cite{okada05}, which contains the best data sample so far. Note that
their analysis, based on Ref.~\cite{bklst}, was done using a simple model for $\phi^h_1$ and they didn't introduce the full expressions for the Coulomb amplitudes $\phi_i^C$, as we do here for consistency.\\
Before going to the conclusion, it is worth mentioning some very recent data at $\sqrt{s}$ = 6.7 GeV \cite{okada07}, 
with a statistically limited accuracy, which might indicate the existence of a non-zero $\phi^h_5$. However
this energy is too low to allow a simple theoretical interpretation of $\phi^h_5$ in terms of a non-zero
Pomeron flip coupling and would require a more elaborated phenomenological analysis,
including dominant Regge contributions.

%%%%%%%%%%%%%%%%%%%%%%%%%%%%
\section{Conclusion}
We have shown, in the context of the impact picture, that the analyzing 
power $A_N$ can be described in the CNI region
by the interference between the non-flip hadronic
amplitude and the single-flip Coulomb amplitude. Unfortunately the data set 
at $\sqrt{s} = 200~\mbox{GeV}$ is too limited to confirm the predicted trend. It
should be extended to make sure this method is a reliable high-energy
polarimeter. The RHIC machine offers a unique opportunity
to measure single- and double-spin observables, with both 
longitudinal and transverse spin directions, and we believe it worthwhile to improve
such measurements, in particular in the small momentum transfer region, as
discussed in Ref.~\cite{bklst}. So far $A_{NN}$ was found consistent
with zero within 1.5$\sigma$ \cite{okada07, bult07}. It is a trivial statement to say that at the moment
we know almost nothing on the $pp$ spin-flip amplitudes at high energy, due to 
the scarcity of previous
experiments performed at CERN and Fermilab. They don't allow us
to make a reliable amplitude analysis, which requires these new measured observables, in a significant
range of momentum transfer. This will be important for our
understanding of spin-dependent scattering dynamics. 
\newpage
\begin{acknowledgments}
We are grateful to Margaret Owens for a careful reading of the manuscript.
J.S. is glad to thank G. Bunce, H. Okada and N. Saito for useful discussions at Brookhaven National
Laboratory.
The work of one of us (T.T.W.) was supported in part by the US Department 
of Energy under Grant DE-FG02-84ER40158; he is also grateful for 
hospitality at the CERN Theoretical Physics Division.
\end{acknowledgments}

\end{document}